
%
\input harvmac
\pretolerance=10000
\def\mod{{\rm mod}}
\def\ket#1{| #1 \rangle}
\def\bra#1{\langle #1 |}
\def\ker{{\rm ker}~}
\def\coker{{\rm coker}~}

\Title{HWS-94/15}
{\vbox{\centerline{Supersymmetry, Vacuum Statistics, and}
  \vskip2pt\centerline{the Fundamental Theorem of Algebra}}}

\centerline{Donald Spector\footnote{$^\dagger$}
{spector@hws.edu}}
\bigskip\centerline{Department of Physics, Eaton Hall}
\centerline{Hobart and William Smith Colleges}
\centerline{Geneva, NY 14456 USA}

\vskip .3in
I give an interpretation
of the fundamental theorem of algebra based on
supersymmetry and the Witten index.  The argument gives a physical
explanation of why a real polynomial of degree $n$ need not have
$n$ real zeroes, while a complex polynomial of degree $n$
must have $n$ complex zeroes.
This paper also addresses in a general
and model-independent way the statistics
of the perturbative ground states (the states which correspond to
classical vacua) in supersymmetric theories with complex and with
real superfields.

\Date{8/94; rev. 6/95}

\nref\fthma{Ahlfors, Lars V.: {Complex Analysis: An Introduction
to the Theory of Analytic Functions of One Complex Variable.}
New York: McGraw-Hill 1966}
\nref\indthm{Alvarez-Gaum\'e, L.: {Supersymmetry and
the Atiyah-Singer Index Theorem.} {Commun. Math.
Phys.} {\bf 90}, 161-173 (1983)}
\nref\smobius{Spector, D.: {Supersymmetry and the
M\" obius Inversion Function.} {Commun. Math.
Phys.} {\bf 127}, 239-252 (1990)}
\nref\SQM{Witten, E.: {Dynamical Breaking of
Supersymmetry.} {Nucl. Phys.} {\bf B188}, 513-554 (1981)}
\nref\WI{Witten, E.: {Constraints on Supersymmetry Breaking.}
{Nucl. Phys.} {\bf B202}, 253-316 (1982)}
\nref\TFT{Witten, E.: {Topological Quantum Field Theory.}
{Commun. Math. Phys.} {\bf 117}, 353-386 (1988)}


Supersymmetry provides some of the richest insights into the
connections between physics and mathematics, with the Witten
index \WI\ serving as one of the
central tools in forging such connections.
Perhaps what is most striking is the range of the
applications of supersymmetry to mathematics; supersymmetry
has been used to prove the Atiyah-Singer
index theorem \indthm, to compute the
topological invariants of manifolds \WI \TFT,
and to derive a variety of results in arithmetic
number theory \smobius.
The central role of the Witten index in these and in many other
physical and mathematical applications stems from the invariance
of the index under deformations of the parameters of a theory.
This makes the index a powerful tool.  It means that the index
may be calculated reliably by simple means, as one need only find
one point in parameter space where it is easily calculable to know
its value at all points in parameter space; this in turn makes
possible the derivation of exact,
non-perturbative results about physical theories and the
mathematical structures they describe (subject to certain
caveats I mention below).

In this paper, using arguments that are not mathematically rigorous
but which are nonetheless instructive and compelling, I extend
the scope of the connections between
supersymmetric physics and mathematical results by showing how
one can use supersymmetry to obtain the fundamental theorem of algebra.
In fact, I will use supersymmetry not only to show
that an $n^{th}$-degree polynomial
over the complex numbers always has $n$ roots, but also to
demonstrate that an $n^{th}$-degree polynomial over the reals
has an even or odd number of real roots, according to whether
$n$ is even or odd, respectively.
(Here, and throughout this paper, I always include multiplicities
when I refer to the number of roots of a polynomial.)
Furthermore, our results will
provide a physical interpretation of why the zeroes in the
complex and real cases behave differently, of why the fundamental
theorem of algebra holds for complex but not real polynomials.
The central piece of the argument is to
associate the zeroes of a polynomial with the classical vacuum
states of a supersymmetric quantum theory, followed by the
use of the Witten index to understand how the number of
such states may or may not change as one changes the parameters
of the theory, and hence how the number of zeroes
may or may not change as the coefficients of the relevant
polynomial are changed.  The polynomial in question is the first derivative
of the superpotential.  On the physical side, our work establishes
some general results regarding the statistics associated with these
classical vacua, and why the statistics of these vacua
exhibit different relationships in the cases of theories
with complex and real superfields, respectively.

The organization of this paper is as follows.  Following this
introduction, I review some of the fundamental properties of the
Witten index and address some basic facts regarding the zeroes
of polynomials.  Then I proceed to a discussion of the zeroes of
polynomials over the reals by computing the Witten index of a
quantum theory with real superfields.  I find that the classical
vacua alternate between bosonic and fermionic statistics, leading
to the conclusion that the number of real zeroes of a real polynomial
and its degree are equal in ${\rm mod }~2$ arithmetic.  Then I
move on to a discussion of complex polynomials, and find that
in the corresponding quantum theories, all the classical vacuum
states have the same statistics.  This, in turn, leads to the
fundamental theorem of algebra, namely that an $n^{th}$-degree
polynomial over the complex numbers has $n$ complex
roots \fthma.  In both cases,
I show how the appearance mathematically of multiple roots
corresponds physically
to the existence of classical vacua with vanishing
perturbative mass gap (i.e., minima with vanishing quadratic
contribution to the Taylor series), and thus develop a physical
understanding of how to identify and treat multiple roots.

In an appendix, I demonstrate that the invariance
of the index under parameter deformations can be established
without first diagonalizing the Hamiltonian.
Since some treatments of diagonalization implicitly
employ the fundamental
theorem of algebra, this technical detail is important; it ensures
us that the argument of this paper is not circular, i.e., that we
are not invoking results that themselves depend on the fundamental
theorem of algebra.

Let me emphasize that the point of this paper is to explore some
intriguing connections between the Witten index of supersymmetric
theories and the fundamental theorem of algebra, obtaining insights
both into the properties of the perturbative zero-energy states
of a supersymmetric theory and into the behavior of zeroes of
polynomials.  As such, I have intentionally focused on the physical
theories that offer the most instructive insights.  While this
leads to a variety of interesting results,
the derivation of these findings
is not entirely rigorous from a mathematical point of view.  The
gaps in rigor stem primarily from two points: the use of field theories
which have never been shown to exist formally, and the use of
perturbative calculations of the index (equivalently, the use of
the ultralocal limit in the functional integral to calculate the
index), a method which is reasonable but which has not been rigorously
proven to be valid in general.  I will return to these points in the
next section.

It is worth pointing out here that there are at least three ways to view
the material presented in this paper.  One can view this paper as
a work of interpretation, recognizing a common mathematical structure
at work in the behavior of the zeroes of polynomials and of the Witten
index of supersymmetric theories.  Such interpretation enriches our
understanding of both fields, and it is this richer understanding,
even in the absence of mathematical rigor, that the reader is
urged to get from this paper.  Second, one can view this paper as
an outline of a proper derivation of the fundamental theorem of
algebra; where there are gaps in rigor is clear, but how to fill in
these gaps, converting an informal argument
into a formal derivation, is an open task.
Third, one can
think of reversing the arguments presented in this paper.  Since one
knows that the fundamental theorem of algebra has been proven rigorously
by other means, the consistency of our analysis of physical theories
(e.g., the perturbative calculation of the Witten index) with the
fundamental theorem of algebra provides a non-trivial (although
obviously not definitive) check on those physical methods, a necessary
but not sufficient test that those physical methods must pass.  I have
thought of this paper primarily in the spirit of the first approach,
but all three approaches have relevance.

\newsec{Index Basics}

In a supersymmetric quantum theory, the Hamiltonian $H$ is
given by the square of the supercharge $Q$,
a Hermitian fermionic operator.  This implies that there
are no states of negative energy.  Furthermore, the operator $(-1)^F$
that measures fermion number anticommutes with the supercharge.
This means that the states of positive energy come in
degenerate bose--fermi pairs.
Consequently, if we define the Witten index as $\tr (-1)^F e^{-\beta H}$
(often written more loosely as $\tr(-1)^F$), we see that
the index calculates the difference between the number of
bosonic and the number of fermionic zero energy states.

Because of the pairing of positive energy states, changes
in the parameters of the theory that maintain supersymmetry
can cause states to
enter or leave the kernel of the supercharge (which is
also the kernel of the Hamiltonian) only in bose--fermi pairs.
Thus, under continuous deformations in the parameters of a
theory (as long as these do not change the behavior of the
potential at infinite field strength or otherwise similarly
change the Hilbert space of the theory \WI),
the index cannot change; it may be calculated at every point
in parameter space by calculating it at one convenient point in
parameter space.  Furthermore, any approximation scheme that
respects supersymmetry will give a correct and exact (not
approximate) value for the Witten index.  This is because the
higher order corrections can only have the effect of
moving states into and out of
the kernel of the Hamiltonian in bose--fermi pairs.

The Witten index is therefore topological in nature, and
so is typically exactly calculable through rather
simple methods.  For our purposes, we will generally calculate
the index using perturbative methods.  Restricting to finite
volume with periodic boundary conditions, we will identify the
classical ground states.  We will then study the perturbative
spectrum about each such state to find its contribution to the
index.  Summing these contributions from all the classical ground
states will then give the index.  One important feature of this
perturbative method is that the perturbative
contribution from the expansion about each ground
state depends only on the properties of the potential
in the neighborhood of that minimum of the potential;
the other vacua have no effect on the index
contribution.  Tunneling between vacua is a higher order
correction, and so does not change the value of the index.
Note that in the functional language, this method of expanding
about the perturbative zeroes is
essentially equivalent to taking the ultralocal
limit, in which the fields are taken to be constant, and hence the
functional integral becomes an ordinary integral.

The heart of this paper lies in the application of this method, and
as such it is important to understand the level of rigor of this method.
Indeed, one way to present this paper is simply to say that it is
a derivation of the fundamental theorem of algebra based on the
assumption that the various supersymmetric physical theories
employed exist and
that perturbative (equivalently, ultralocal) calculation of the
Witten index of such theories is valid.
Of course, in the case of the supersymmetric
field theories we use, it is not established that such theories
exist rigorously from a mathematical point of view.  And, even if they do
exist, the validity of the perturbative or ultralocal approximation
for calculating the index is still an open question --- one can argue
that such an approximation is plausible via Fourier expansions,
for example, but such arguments are formal, relying on such requirements
as various
functional integrals being well-defined, various limits being
non-singular and commuting with other operations, etc.  In the
case of non-relativistic quantum mechanics, the perturbative or ultralocal
approximation rests on a better footing, and although there is no
general proof of its applicability, at least in particular theores,
with care, one can show its validity, although proper treatment of this
requires careful analytical work (e.g., showing that the Hilbert
space of states does not change as one includes perturbative and
non-perturbative corrections).  And, obviously,
proving rigorously the validity of these
methods in the case of field theories is an open question
that will not be answered in the near future, barring dramatic progress
in constructive field theory.

Note that if I removed all references to field theories in this paper,
by speaking only of non-relativistic quantum mechanics and its
complex generalization, this would enhance the rigor of the argument.
I have nonetheless opted against this choice, favoring instead to
seek greater insight into the interplay of physics and mathematics
that occurs in supersymmetric theories.
By using field theories, I am able to explore
the behavior of perturbative ground states of
supersymmetric theories, and to use this behavior to understand
the behavior of the zeroes of polynomials, provided of course the
physical theories
in question (or at least very similar theories) exist.
After all,  my main goal is to use supersymmetry to enhance
the understanding of the fundamental theorem of algebra.
As we know, technically sound proofs of the fundamental theorem of
algebra already exist; but new insights into why the theorem is true
nonetheless further our understanding of the physical and mathematical
structures in question.  Thus the reader is urged to remember
throughout this paper that the arguments presented are valuable for
the insights they provide and compelling for the structures they
suggest, but that the arguments from which these results
are obtained are not mathematically rigorous.

Having set this context, let us return to a consideration of the
perturbative calculation of the index.
Suppose one considers a theory with superpotential $W(\phi)$. Let
$P(\phi) = {\partial W(\phi) \over \partial \phi}$.
The supersymmetric vacuum states classically are given
by the zeroes of $P(\phi)$ since the scalar potential is
$V(\phi) = |P(\phi)|^2$.
These zeroes fall into two categories.  Either the perturbative
excitation spectrum around the zero has no massless particles
or it does have massless particles.  If the classical vacuum has
no massless particles in its excitation spectrum, this
classical vacuum contributes either $+1$ or $-1$ to the
Witten index.  If the classical zero energy state does have massless
particles in its perturbative spectrum, the situation
needs to be studied more carefully.  It is the connection between
the Witten index and the zeroes of $P(\phi)$ that will form
the basis of our interpretation of the fundamental theorem of algebra.
By choosing $P(\phi)$ to be a polynomial of degree $n$, we can
address the questions we wish to address, establishing results
regarding the number of zeroes of an $n^{th}$-degree polynomial.

Note that any two $n^{th}$-degree polynomials $P(\phi)$
must produce the same Witten index (modulo the possibility
in the real case for two polynomials to produce indices of the
same magnitude but opposite sign; I discuss this in the next
section).
Changing the sub-leading coefficients in $P(\phi)$ does not
change the asymptotic behavior of the potential, and hence does
not change the index.  Likewise, changing
the leading coefficient does not change the index, as long
as this leading coefficient is not made to vanish.
We will use the equality of the index associated with
all $n^{th}$-degree polynomials $P(\phi)$
within particular classes of theories (keeping track of the
possible sign flip mentioned above) to relate
the number of zeroes of various $n^{th}$-degree polynomials.

Finally, it is worth making the rather obvious remark that
an $n^{th}$-degree polynomial cannot have more than
$n$ roots.  The easiest way to see this is by observing first that
if $\phi = \phi_0$ is a solution to $P(\phi)=0$,
then $\phi -\phi_0$ is
a factor of $P(\phi)$ (which is obvious once one shifts variables
by $\tilde \phi = \phi - \phi_0$), and observing that
an $n^{th}$-degree polynomial cannot have more than $n$
linear factors.

\newsec{Polynomials over the Reals}

In this section, we consider the use of the index to study the zeroes
of polynomials over the reals.  We will first discuss this case
in general terms, and then look at our results in the context
of a specific model.  We will use the language of field theory in
the general discussion because of the useful insights one obtains,
and then move on to the mathematically
less precarious case of non-relativistic
quantum mechanics when we turn to a specific model.

Let us consider a supersymmetric theory in which
the superpotential $W(\phi)$ is a real-valued function of the
real-valued field $\phi$, as occurs, for example, in
supersymmetric quantum mechanics and in $2+1$ dimensional
supersymmetric scalar field theory.
As in the previous section, define $P(\phi) = {\partial W
\over \partial \phi}$.  To calculate the index using perturbative
methods, one
must identify the zeroes of $P(\phi)$, since the potential
is $V(\phi) = \left(P(\phi)\right)^2$.  We are interested in the
case that $P(\phi)$ is a polynomial in $\phi$.

We now wish to find the index for a theory in which $P(\phi)$ is an
$n^{th}$-degree polynomial.
Since all polynomials of a given degree have the same index, we proceed
by first considering a representative polynomial of given degree $n$;
from this, we will learn the index associated with and gain
insight into the number of zeroes of
any $n^{th}$-degree polynomial $P(\phi)$.  It turns out that we
only need to consider two representative polynomials,
one of even degree, and one of odd degree.

First, consider as a representative even degree polynomial
$P(\phi) = \phi^n+1$, where $n$ is even.  This polynomial
manifestly has no real roots.  As a consequence,
the Witten index for the corresponding theory (in which
$W(\phi)=\int P(\phi) d\phi$ is an odd degree polynomial) is zero.
 From this we can conclude that the Witten index of {\it any}
theory for which $P(\phi)$ is an even degree polynomial is zero.
Ignoring for the moment the situation when
there are zeroes of $P(\phi)$ about which there
is no mass gap, we see that, since each zero of $P(\phi)$
generically contributes
either $+1$ or $-1$ to the Witten index, $P(\phi)$ must have
an even number of zeroes (an equal number of bosonic and fermionic
ones) when $P(\phi)$ is a real polynomial of even degree.

Now, as our representative odd degree polynomial, consider
\hbox{$P(\phi) = (\phi+1)(\phi^{n-1}+1)$}, where $n$ is odd.  This
polynomial manifestly has only one real root, about which
the perturbative spectrum has a mass gap.  As a consequence,
the Witten index for the corresponding theory is either $+1$
or $-1$.  (The actual sign does not matter for our purposes.)
This means that any odd degree polynomial has Witten index $\pm 1$.
Hence (excepting for the moment the situation in which
there are zeroes with no mass gap), an odd degree polynomial has an
odd number or zeroes.

Note, incidentally, that the statement that any $n^{th}$-degree
polynomial can be deformed continuously into any other
$n^{th}$-degree polynomial comes with
a caveat.  Remember that as long
as the leading asymptotic behavior of the potential does not change
(keeping a canonical kinetic term throughout, of course),
the index does not change.  Now for a real polynomial, the leading
coefficient cannot change from positive to negative without passing
through zero, at which point the leading behavior of the potential
is different.  Thus it is possible that an $n^{th}$-degree polynomial
will produce a different index depending on whether its leading
coefficient is positive or negative.  However,
modifying our examples above by multiplying each representative
polynomial by $-1$, we see that the only possible difference in
the index value in the case of positive versus negative leading
coefficient is a difference in sign.  For even polynomials, this
leaves the index as zero; for odd polynomials, we again conclude
that the index is $\pm 1$.  Thus this possible change in sign, while
relevant in other contexts,
has no actual bearing on the arguments presented in his paper,
and so we will not pursue it further here.

What happens if there are classical vacua for which
there is no mass gap in the perturbative spectrum?  Performing a Taylor
expansion about some root, say $\phi_0$, tells us that in the
neighborhood of such a root, $P(\phi) \approx C (\phi -\phi_0)^r$.
This is sufficient for studying the contribution
to the index from the perturbative expansion about the classical
vacuum point $\phi = \phi_0$.
Note that if $r=1$, there is a mass gap, and we have the case
considered previously; if $r>1$, there is no perturbative
mass gap.  At this point, the system
looks like one in which $\partial W/\partial \phi$ is an
$r^{th}$-degree polynomial, and so we see that the contribution
to the index from the perturbative expansion about this
vacuum state is simply $r~\mod ~2$.
Also, as is easily seen by shifting variables to
$\tilde \phi = \phi - \phi_0$, if a polynomial
$P(\phi) \approx C (\phi -\phi_0)^r$ in the neighborhood of
$\phi = \phi_0$, then $(\phi -\phi_0)^r$ is a factor of $P(\phi)$.

This means that if we introduce the familiar notion of
a multiple zero (i.e., counting it as $r$ zeroes of $P(\phi)$ if
$(\phi - \phi_0)^r$ is  factor of $P(\phi)$), our previous statements
regarding the number of zeroes of $P$ remain unchanged.
Since an $r$-fold zero contributes $\pm (r~\mod~2)$ to the index,
in order for a polynomial of degree $n$ to produce the
required index, namely $n~\mod~2$, that polynomial must have
an even or odd number of real zeroes, respectively, according to
whether $n$ is even or odd,
provided we count multiplicities when we enumerate the roots.
(Of course, the total
number of zeroes --- indeed, of linear factors --- can never exceed
the degree of the polynomial.)

It is worth noting that our result for real polynomials says
that, as the parameters of a real polynomial change,
the number of zeroes of that polynomial can only change
by two (or a multiple of two)
at a time, so as to preserve the value of the index.  In
the physical interpretation of this result, this is the statement
that classical vacua can only appear or disappear in bose--fermi
pairs.  There is a standard algebraic interpretation of this result, too.
We can view a real polynomial as a special example of a complex
polynomial (which must, as we will see below, have $n$ complex
roots).  Since the complex roots of a real polynomial must occur
in complex conjugate pairs, we have recovered physically the
familiar algebraic result that the only way for the number
of real roots of a real polynomial to change is for its complex
roots to appear or disappear in complex conjugate pairs.

As an explicit example of our construction for real polynomials,
we can consider supersymmetric quantum mechanics in one spatial
dimension \SQM.
Including the $\hbar$'s, the Hamiltonian is given by
\eqn\sqmham{H =
-{\hbar^2 \over 2m}{d^2 \over dx^2} + P(x)^2
      + \sigma_3{\hbar\over\sqrt{2m}} {dP(x)\over dx}~~~.}
The operator $(-1)^F$ is simply $\sigma_3$.  The classical zero
energy states are given by the solutions of
\eqn\realpoly{P(x)=0~.}
Some of these solutions may be bosonic and some fermionic, depending
on whether they are eigenstates of $\sigma_3$ with eigenvalue $+1$
or $-1$, respectively.
As Witten showed, the exact quantum theory has either no zero energy
states or exactly one zero energy state.  Thus, the index is either
$0$ or $\pm 1$, respectively, in accord with what we found based on
general arguments;  in fact, when $P(x)$ goes as
$x^n$, Witten showed in \SQM\ that there is no zero energy state
when $n$ is even, and that there is one zero energy state when $n$
is odd.

In perturbation theory, these index results arise by expanding
about each point where $P(x)$ vanishes.
Suppose $x_0$ corresponds to a zero of $P(x)$ and hence to a classical
vacuum state.  The Hamiltonian near this point to leading order
is
\eqn\approxham{H \approx
-{\hbar^2\over 2m}{d^2 \over dx^2}
    +\bigl(  (x-x_0) P^\prime(x_0) \bigr)^2
    +\sigma_3 {\hbar\over\sqrt{2m}} P^\prime(x_0)  ~~~.}
This is a harmonic oscillator potential; calculating, we see that if
$P^\prime(x_0)$ is positive (respectively, negative), this Hamiltonian
has a single fermionic (respectively, bosonic) zero energy state.
Hence the classical vacuum contributes an amount to the index
equal to $-sign(P^\prime(x_0))$.
Thus, depending on this sign, the classical vacuum is associated
with either a bosonic or fermionic perturbative zero energy state.
Since the fermion number is given by minus the sign of $P(x_0)$, it
therefore follows that, as one proceeds along the spatial axis,
the vacua one encounters are alternately bosonic and fermionic.
(If $P(x)$ has vanishing slope
at the classical vacuum point, we simply need to refine the argument
to include the notion of multiple zeroes, just as we did earlier.
There is little to be gained by doing that here, so we leave it as
an exercise for any interested readers.)

Incidentally, this alternation of the fermion number of the classical
vacua can be derived from more abstract index arguments.  Since we
can deform the location of the zeroes of the superpotential without
changing the index, we can choose all but a pair of zeroes to be
very far from each other, so that the physical effect of those faraway
zeroes is negligible.  Perturbatively, then, this theory looks just
like a theory in which $P(x)$ is quadratic, plus the contributions
of the faraway zeroes.
Since the index in a theory with quadratic $P(x)$ is zero, the
statistics of these two nearby zeroes must cancel each
other.
Any pair of adjacent zeroes of $P(x)$ can be isolated in the way
just described above,
but the order of the zeroes of $P(x)$ cannot change under smooth,
non-singular deformations.
Thus any pair of adjacent
zeroes of $P(x)$
must have opposite fermion number, which in turn means that the
classical vacua correspond, in alternating fashion, to bosonic and
fermionic states.
Note, too, that this gives us another way to understand the
contribution to the index at a root of $P(\phi)$ which is also
a point of inflection.  We can model the perturbative situation
of $P(x) \approx C (x-x_0)^r$ as the coalescence
of $r$ adjacent zeroes, all deformed to the point
$x=x_0$.  If $r$ is even, then this coalescence will always involve an
equal number of bosonic and fermionic vacua (since the fermion
number alternates), producing index $0$; if $r$ is odd, by a similar
argument, we see that this coalescence will produce index $\pm 1$ about
this classical vacuum.

\newsec{Polynomials over the Complex Numbers}

The consideration of the zeroes of polynomials over
the complex numbers proceeds in a way similar to what we
have already done for polynomials over the reals, although the results
are quite different.  In short, we will see that every
$n^{th}$-degree complex polynomial has $n$ complex roots.
In this section, I will use the language of field theory; the reader
has already been alerted to the ways in which this undercuts the
rigor of the argument, but we expect that the reader will nonetheless
find the results instructive and insightful.

Let us consider a supersymmetric theory with complex superfields,
such as supersymmetric scalar field theory in $3+1$ dimensions.
As in the real case, we define $P(\phi) = {\partial W \over
\partial \phi}$, where the superpotential $W(\phi)$ is now a
complex-valued function of the complex field $\phi$.
The potential is
$V(\phi)=|P(\phi)|^2$.  The perturbative computation of the index
can thus be achieved by determining all solutions of $P(\phi)=0$,
and then expanding about each of these.
Note that any two $n^{th}$-degree complex polynomials $P(\phi)$
will produce the same value for the index.  This result is
slightly stronger than in the real case.  In the complex case, the
leading coefficient of $P(\phi)$ can be changed continuously from a
positive to a negative number without ever passing through zero.
Thus all $n^{th}$-degree polynomials yield exactly the same index,
with the same magnitude {\it and} sign.
This contrasts with the real case
in which changing the sign of the leading coefficient could change
the sign, but not the magnitude, of the index.

One additional
simplification in the complex case is that we will not need to
distinguish between even and odd degree polynomials, as will readily
become apparent.

To proceed, then, we first pick a representative $n^{th}$-degree
complex polynomial and calculate the index associated with it.  We then
will use this result to find the index associated with any
$n^{th}$-degree polynomial, which we will then use to show that
an arbitrary $n^{th}$-degree polynomial has exactly $n$ roots.

As our representative $n^{th}$-degree polynomial, let us consider
\eqn\crep{P_0(\phi)=(\phi -c_1)(\phi -c_2)\cdots (\phi - c_n)~~~,}
with the complex constants $c_j$ all distinct from each other.
Clearly, this polynomial has $n$ complex roots, with non-zero mass gap
in the perturbative spectrum about each of these classical
vacua.  We must now determine which
of these classical vacua correspond to bosonic states and which
to fermionic states, so that we can compute the index.

Under continuous non-singular deformations of the parameters
of the theory, the statistics of each individual vacuum cannot change,
as the eigenvalues of $(-1)^F$ can take on only the discrete
values $+1$ and $-1$.  Let us concentrate for
the moment on two of the vacua, say $\phi = c_1$ and $\phi = c_2$,
as we undertake certain deformations in the parameters of the theory.

Note that one can
vary $\phi -c_2$ smoothly to $\phi -c_1$, while
varying $\phi -c_1$ smoothly to $\phi -c_2$.  To do this,
consider
\eqn\Psigma{\eqalign{
P_\sigma(\phi)=
     (\phi -&[{c_1+c_2 \over 2} + {c_1-c_2\over 2}e^{i\pi \sigma}])
      (\phi -[{c_1+c_2 \over 2} - {c_1-c_2\over 2}e^{-i\pi \sigma}])
      \cr
      &\times (\phi - c_3)\cdots (\phi - c_n)
}}
As $\sigma$ varies from $0$ to $1$, $P_\sigma$ continuously deforms
so that the vacua at $\phi=c_2$ and $\phi=c_1$ smoothly switch
locations.  Thus, by the fixed value of the statistics
of each vacuum individually under continuous changes of the
parameters of a theory,
the statistics at $\phi=c_1$ when $\sigma = 0$ must
be identical to the statistics at $\phi=c_2$ when $\sigma=1$; and
the statistics at $\phi=c_2$ when $\sigma = 0$ must
be identical to the statistics at $\phi=c_1$ when $\sigma=1$.
On the other hand, the polynomial is exactly the
same at $\sigma=0$ and $\sigma=1$.
Thus the statistics of the vacuum at
$\phi = c_1$ is the same whether $\sigma=0$ or $\sigma=1$,
and the statistics of the vacuum at $\phi = c_2$ is
the same whether $\sigma=0$ or $\sigma=1$.
Putting this all together, we see that in our original polynomial,
the vacua at  $\phi=c_1$ and  $\phi=c_2$ must have the same statistics.

Now there was nothing special about these two particular vacua.
Hence we see that all the classical vacuum solutions for this polynomial
have the same fermion number, and thus the Witten index
is $\pm n$.
(We do not need to determine the sign.)

So, in this particular theory, the
index has value $\pm n$.  This is a theory, however, in which
$P(\phi)$ is an $n^{th}$-degree polynomial.  As we have argued
above, any two theories in which $P(\phi)$ is an $n^{th}$-degree
polynomial over the complex numbers must produce the same index.
This means that any other theory in which $P(\phi)$ is an
$n^{th}$-degree polynomial must have a Witten index of
value $\pm n$, whether we know how to write the polynomial in the
factorized form in \crep\ or not.

What are the further implications of this result for general
$n^{th}$-degree polynomials $P(\phi)$?
Since the index for such a theory is $\pm n$, and
since generically the zeroes of $P(\phi)$ will all exhibit
mass gaps in their respective perturbative spectra,
we see that a generic $n^{th}$-degree
polynomial must have $n$ zeroes.  Since achieving an index
of $\pm n$ requires at least $n$ zeroes, and since an
$n^{th}$-degree polynomial can have at most $n$ zeroes, we
can conclude that an $n^{th}$-degree complex polynomial has
exactly $n$ zeroes.

Note further that
since the index is $\pm n$ and all the classical vacua where
$P^\prime(\phi)\ne 0$ can be deformed into each other and
hence have the same statistics, we can rule out on index grounds
alone the possibility that there are more than $n$ zeroes
for an $n^{th}$-degree polynomial $P(\phi)$.  (In fact, using
the observations developed below regarding zeroes which are also
points of inflection, one can extend this to use index arguments
alone to show that there are no more than $n$ zeroes even when
some are located at points where $P^\prime(\phi)$ vanishes.)

What if there are zeroes of $P(\phi)$ about which there is
perturbatively no mass gap (i.e., the first term in the Taylor series
for the potential about that point is higher than quadratic)?
What happens at such a point?

As in the real case, we can understand such a point independent of
what is happening elsewhere in the potential.  If the polynomial
$P(\phi) \approx C (\phi - c)^r$ to lowest order near the
zero $\phi = c$, then, calculating the contribution to the index
from this point perturbatively, we see that the contribution to
the index at this point is the index associated with an
$r^{th}$-degree polynomial.
Consequently, such a point can contribute an amount to the index of
$\pm r$, only.  Since we have ultimately to reach a total index of
$n$, and the polynomial can have no more than $n$ linear factors,
one can infer that the sign of this index contribution must
be the same as the sign of all the other index contributions
from all the other perturbative vacua.
(Recall that for a polynomial, if $P(\phi) \approx C (\phi - c)^r$
near $\phi = c$, then $(\phi - c)^r$ is a factor of $P(\phi)$.)

Alternatively, we can see that the sign is the same directly.  We
can obtain a function which is perturbatively identical to the
$C (\phi - c)^r$ considered above by considering the coalescence
of $r$ vacua which have mass gaps.  Since all the vacua with mass gaps
contribute to the index with the same sign
(the same not only among the
$r$ vacua that are coalescing, but among all the vacua),
the coalescence of $r$ such vacua must give a contribution with
the same sign as all these other vacua, too (although larger in
magnitude by a factor or $r$).

Thus, if we introduce the familiar notion of the multiplicity of a
zero, we have just concluded, subject to the limitations of rigor
pointed out above, that all complex $n^{th}$-degree
polynomials
have $n$ complex roots.  What the work above demonstrates is that
an $r^{th}$-degree zero contributes an amount $r$ to the index;
all the zeroes contribute to the index with the same sign; and
the total index is $n$ for a polynomial of degree $n$.  This means
that, counting multiplicities, a complex polynomial of degree $n$ has
exactly $n$ roots. This is the fundamental theorem of algebra.
Our argument rested on our determination that
any  $n^{th}$-degree polynomial $P(\phi)$ leads to
a theory with the same value for the
Witten index, namely $\pm n$, which in turn implies that
$P(\phi)$ has $n$ linear factors.

Note that we have found some essential differences
between the complex and real cases.  The most significant
is that in the complex case, the locations of the vacua in
field space could be interchanged smoothly, without colliding,
by a continuous change in the parameters of the theory.
This is what led to an index of value $\pm n$,
which in turn led to the fundamental theorem of
algebra by forcing the number of zeroes to be $n$.
In the real case, this smooth
type of interchange is not possible, which makes it
possible for the vacua not all to have the same fermion number.
Indeed, this occurs; as we have seen,
the real case always has vacua of alternating fermion number.

Before closing, it is worth noting that one approach to showing
that a complex polynomial of degree $n$ has $n$ complex roots
is first to show that any complex polynomial has at least
one complex root.  From this, one can proceed by induction to show
that an $n^{th}$-degree polynomial has $n$ roots, by pulling out
a linear factor for the first root one knows about, and then
studying the polynomial of degree $n-1$ this leaves behind.
We could have used the above index arguments to recognize
immediately that, since any complex polynomial is associated with
non-zero index, it must have at least one zero, in order for the
perturbative index calculation to yield a non-zero result.
One could then proceed via induction to show mathematically that
such a polynomial has $n$ complex roots.
Our goal here, however, has been to show
the rich set of relationships that exist between
the physical properties of supersymmetric quantum
theories and the mathematical results regarding the zeroes of polynomials,
and thus we have chosen instead to seek an understanding of
the broader behavior of these zeroes from a physical perspective.

\bigbreak\bigskip\bigskip\centerline{{\bf Acknowledgments}}\nobreak
I thank the LNS Theory Group at Cornell University, where
parts of this research were conducted, for its hospitality.
This work was
supported in part by NSF Grant. No. PHY-9207859.

\appendix{A}{Invariance of the Index}

In this appendix, I derive the topological invariance of the
Witten index
\eqn\widef{\Delta = \tr (-1)^F e^{-\beta H} ~,}
by which I mean that I show the index to be unchanged under
parameter deformations and to be calculable exactly using
approximation methods.  The reason for doing this is that typically
discussions of the invariance of the index invoke
diagonalization of the Hamiltonian
along the way, and some treatments of the diagonalization of
operators implicitly use the fundamental theorem
of algebra (invoked for finite dimensional matrices, followed
by the taking of an appropriate limit),
something we must avoid if
we are to use the invariance of the Witten index to derive
the fundamental theorem of algebra.  Thus, in this appendix,
I establish explicitly that the fundamental theorem of algebra is
not necessary to derive the topological invariance of the index.

Let the Hamiltonian $H$ and the supercharge $Q$ be Hermitian
operators related by $H = Q^2$.  They act on a Hilbert space
$\cal H$.  There is also an operator $(-1)^F$ which anticommutes
with $Q$ and squares to the identity.  Note that the operator
$H$ is positive semi-definite, and so in any state, the
expectation value of $H$ is non-negative, and it can only be
zero for a state in the kernel of $H$.  In addition,
let us let ${\cal H}^\prime = \coker H$;
this is a subspace of $\cal H$.

Note first that $\ker Q = \ker H$.

Second, note that we can choose as a basis for $\cal H$
states which are eigenstates of $(-1)^F$.
To do this, given some basis of kets $\ket{n}$, we define
a new basis using the kets
\eqn\eigenpm{\ket{n} \pm (-1)^F \ket{n} ~,}
which have eigenvalues $\pm 1$, respectively, under $(-1)^F$.

 From now on, I use such a basis.  Let $B$ and $F$ denote, respectively,
the bosonic and fermionic subspaces of $\cal H$,
and let $B^\prime$ and $F^\prime$ denote, respectively,
the bosonic and fermionic subspaces of ${\cal H}^\prime$.

It is straightforward to see that, given $\ket{f^\prime} \in F^\prime$,
$Q\ket{f^\prime} \in B^\prime$.  Clearly,
$Q\ket{f^\prime} \in B$.  Let $\ket{b_s}$ be a bosonic
state in $\ker H$. Then $\bra{b_s}Q\ket{f^\prime} = 0$,
and thus we see that $Q F^\prime \subset B^\prime$.
In an entirely analogous fashion,
we see that $Q B^\prime \subset F^\prime$.

The next step is to show that, in fact, $Q F^\prime =  B^\prime$
and $Q B^\prime = F^\prime$.  This can be obtained in one
of two ways.

Method 1: We start from the existence of an
inverse $H^{-1}$ on ${\cal H}^\prime$.
Then given a bosonic ket $\ket{b^\prime}$, we can
write it in the form $\ket{b^\prime} = Q\ket{f^\prime}$,
by defining the fermionic ket $\ket{f^\prime} = H^{-1}Q\ket{b^\prime}$.
This establishes that $B^\prime \subset QF^\prime$.
Since we have both
$Q F^\prime \subset B^\prime$ (see above)
and $B^\prime \subset QF^\prime$, we see that $Q F^\prime =  B^\prime$.
Likewise, $Q B^\prime = F^\prime$.

Method 2: We start from the (physically
motivated) statement that the range
of $H$ is ${\cal H}^\prime$.
Now we already know that $Q\ket{f^\prime} \in B^\prime$. Applying
$Q$ to this, we find
\eqn\contain{Q^2 F^\prime \subset QB^\prime \subset F^\prime~.}
But in order that the range of the Hamiltonian be the full
${\cal H}^\prime$, we see that $Q^2 F^\prime = F^\prime$.
Thus it follows that $Q B^\prime = F^\prime$ and, likewise,
that $Q F^\prime =  B^\prime$.

Moving on, I now consider
\eqn\tracesep{\tr (-1)^F e^{-\beta H} =
              \tr (-1)^F e^{-\beta H}\Bigr|_{\ker H} +
               \tr (-1)^F e^{-\beta H}\Bigr|_{{\cal H}^\prime}~.}
(Note: If the sum is not absolutely convergent, one should take
the sum in the order implicit in the bose--fermi grouping of states
I use below.)
Now let the kets $\ket{b^\prime_j}$ form a basis for $B^\prime$.
Define
\eqn\fbasis{\ket{f^\prime_j} =
  {Q \ket{b^\prime_j} \over
             \bra{b^\prime_j}H\ket{b^\prime_j}^{1/2}}~.}
Clearly, since $QB^\prime = F^\prime$, the set of vectors
in \fbasis\ spans $F^\prime$.  Do they in fact form a basis
for $F^\prime$?  The answer is that they do.
Otherwise, there would be constants $c_j$ such that
\eqn\sumzero{\sum_j c_j \ket{f^\prime_j} = 0~.}
This would mean then that
\eqn\alsozero{Q\Bigl( \sum_j \alpha_j \ket{b^\prime_j} \Bigr) = 0~,}
(where $\alpha_j = c_j /  \bra{b^\prime_j}H\ket{b^\prime_j}^{1/2}$),
which cannot be, since the $\ket{b^\prime_j}$ form a basis
for ${\cal H}^\prime$, and so a linear combination of them cannot
belong to $\ker H$.

Thus we can set up a one-to-one correspondence between the basis
vectors of $B^\prime$ and the basis vectors of $F^\prime$.
It is a simple exercise to show that each member of a pairing of
bosonic and fermionic basis vectors (as paired above) yields the
same expectation value of $H$ or any function of $H$; that is,
\eqn\evsame{\bra{b^\prime_j}H\ket{b^\prime_j} =
            \bra{f^\prime_j}H\ket{f^\prime_j}~.}
Thus,
\eqn\trvanish{
 \bra{b^\prime_j}(-1)^F H\ket{b^\prime_j} +
   \bra{f^\prime_j}(-1)^F H\ket{f^\prime_j} = 0~,}
Consequently, the Witten index receives a net vanishing contribution
from all the states in ${\cal H}^\prime$, and thus
\eqn\traceker{\tr(-1)^Fe^{-\beta H} = tr (-1)^F\Bigr|_{\ker H}~.}

Note further that we have shown that the basis for ${\cal H}^\prime$
can be organized in bose--fermi pairs with common values for the
expectation value of the Hamiltonian. (We will refer to such
pairs as degenerate because they give the same expectation
value for the Hamiltonian, even though they are not necessarily
eigenstates of the Hamiltonian.) If an approximation scheme
erroneously determines whether a state is in $\ker H$ or
${\cal H}^\prime$, it can do so only in
degenerate bose--fermi pairs, as long
as the approximation scheme respects supersymmetry. Thus using
an approximate method to calculate the Witten index yields an exact
results.  Also, if we deform the parameters of the Hamiltonian, as long
as this leaves the Hilbert space unchanged (which will happen
generically when the asymptotic behavior of the potential is
unchanged), it leaves the Witten index unchanged, as states can
only move between $\ker H$ and ${\cal H}^\prime$ in
degenerate bose--fermi pairs.

Thus we have shown that the Witten index is indeed invariant and is indeed
exactly calculable when treating the system approximately, and
we have shown this
without reference to diagonalizing $H$. This makes clear that the
topological invariance of the Witten index does not depend, explicitly
or implicitly, on the fundamental theorem of algebra.

\listrefs
\bye